\documentclass[aps,prl,amsmath,amssymb,twocolumn,showpacs,floatfix,superscriptaddress]{revtex4-1}
\usepackage{graphicx}
\usepackage[english]{babel}
\usepackage{times}
\usepackage{dcolumn}
\usepackage{units}
\usepackage[usenames]{color} 
\begin{document}

\title{Tunable Kondo effect in double quantum dots coupled to ferromagnetic
contacts}
\author{Rok \v{Z}itko}
\affiliation{Jo\v{z}ef Stefan Institute, Jamova 39, SI-1000 Ljubljana,
Slovenia}
\affiliation{Faculty  of Mathematics and Physics, University of Ljubljana,
Jadranska 19, SI-1000 Ljubljana, Slovenia}
\author{Jong Soo Lim}
\affiliation{Departament de F\'{i}sica, Universitat de les Illes Balears,
  E-07122 Palma de Mallorca, Spain}
 \affiliation{Institut de F\'{i}sica Interdisciplinar i de Sistemes Complexos
  IFISC (CSIC-UIB), E-07122 Palma de Mallorca, Spain}
  \author{Rosa L{\'o}pez}
\affiliation{Departament de F\'{i}sica, Universitat de les Illes Balears,
  E-07122 Palma de Mallorca, Spain}
\affiliation{Institut de F\'{i}sica Interdisciplinar i de Sistemes Complexos
  IFISC (CSIC-UIB), E-07122 Palma de Mallorca, Spain}
\author{Jan Martinek}
\affiliation{Institute of Molecular Physics,
  Polish Academy of Sciences, Smoluchowskiego 17,
  60-179 Pozna\'n, Poland}
\author{Pascal Simon}
\affiliation{Laboratoire de Physique des Solides, CNRS UMR-8502, Univ. Paris Sud, 91405 Orsay Cedex, France}
\date{\today}

\begin{abstract}
We investigate the effects induced by spin polarization in the contacts attached to a serial double quantum dot. The polarization generates effective magnetic fields and suppresses the Kondo effect in each dot. The super-exchange interaction ($J_{\mathrm{AFM}}$), tuned by the inter-dot tunnelling rate $t$, can be used to compensate the effective fields and restore the Kondo resonance when the contact polarizations are aligned. As a consequence, the direction of the spin conductance can be controlled and even reversed using electrostatic gates alone. Furthermore, we study the associated two-impurity Kondo model and show that a ferromagnetic exchange coupling  ($J_{\mathrm{FM}}$) leads to an effective spin-1 exchange-anisotropic Kondo model which exhibits a quantum phase transition in the presence of partially polarized contacts.
%transition between a state with fully screened impurity moment to a
%state with residual entropy.
\end{abstract}

 
% 72.10.Fk Scattering by point defects, dislocations, surfaces, and
% other imperfections (including Kondo effect)
% 72.15.Qm Scattering mechanisms and Kondo effect (see also 75.20.Hr
% Local moments in compounds and alloys; Kondo effect, valence
% fluctuations, heavy fermions in magnetic properties and materials)
% 72.25.Dc Spin polarized transport in semiconductors 
 
\maketitle

\newcommand{\expv}[1]{\langle #1 \rangle}
\newcommand{\vc}[1]{\mathbf{#1}}

%{\it Introduction} -- 
The study of spin-polarized transport in quantum dots (QDs), motivated by potential application for
spintronic devices, has recently drawn a lot of attention both theoretically \cite{PhysRevB.75.165303,PhysRevLett.89.286803,PhysRevB.71.245116,Hamaya:09} 
 and experimentally  \cite{Pasupathy01102004,Calvo09,PhysRevLett.104.036804}. 
 %It has been found that 
 Ferromagnetic contacts affect the dot charge dynamics so that
spin-dependent tunnelling rates renormalize differently the dot 
level for each spin orientation \cite{PhysRevB.75.165303,PhysRevLett.89.286803,Hamaya:09}.
This is reflected in the appearance of  
an effective magnetic field in the dot \cite{PhysRevB.76.045321,PhysRevLett.92.056601,
PhysRevB.72.121302,PhysRevLett.91.247202,PhysRevLett.91.127203}
which suppresses the many-body Kondo state \cite{PhysRevLett.91.247202,PhysRevLett.91.127203}, routinely observed
at low enough temperatures in QDs.
However, this Kondo state can be restored by properly tuning the QD level position ($\epsilon$) via a gate potential
\cite{PhysRevB.72.121302,PhysRevLett.92.056601}. 
Additionally, when the polarizations in the ferromagnetic electrodes
are non-collinear, the induced effective magnetic field depends on the relative orientation of the two easy-axis, but it can, again, be compensated \cite{PhysRevB.75.045310}.

Double quantum dots (DDs) have been extensively studied because they offer interesting perspectives 
in quantum spintronics such as spin-based 
quantum computation \cite{Loss:98}, Pauli spin blockade \cite{Ono:02,Fransson:06}, spin pumping \cite{Platero:05}, etc.
Moreover, DDs offer a natural experimental realization of the two-impurity Kondo problem (2IKP)   
\cite{PhysRevB.40.324,PhysRevB.39.3415,PhysRevLett.68.1046}. 
The interplay between the single dot Kondo effect and the interdot interaction has been extensively studied theoretically \cite{PhysRevLett.82.3508,PhysRevB.63.125327,PhysRevLett.97.166802,Sela:09} 
and its experimental realization has been reported \cite{Jeong21092001,PhysRevLett.92.176801}.
Coupling such DDs to ferromagnetic 
contacts \cite{Tanaka:04,PhysRevB.77.245313} adds a further 
experimental handle to the system, which is distinct from applying an external magnetic field and may have 
interesting applications in spin-dependent transport. 
\begin{figure}[t]
\includegraphics[clip,width=8.cm]{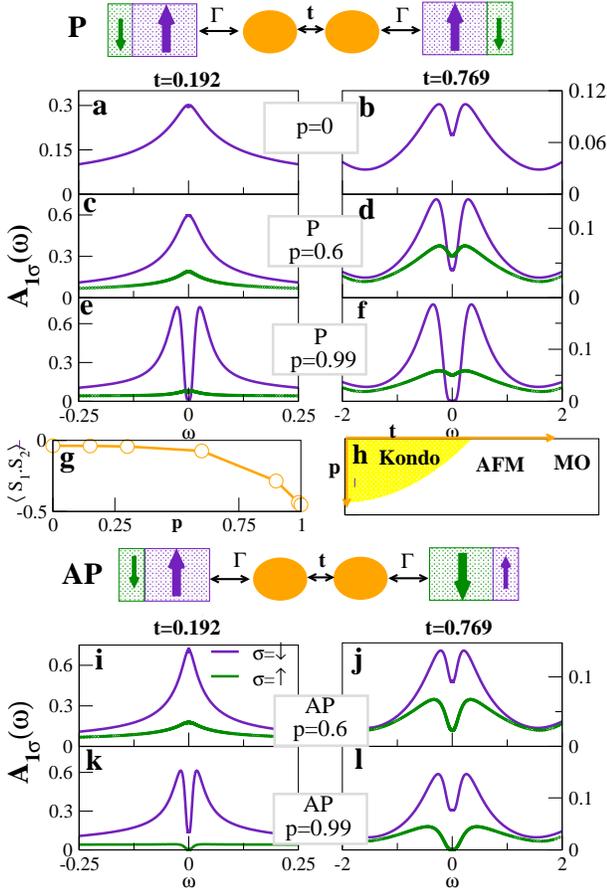}
\caption{(Color online) Upper inset: Illustration of a  serially coupled double dot set-up for
the parallel P configuration [majority(minority) spins of both reservoirs are down(up) spins. $t$ is the 
inter-dot tunneling amplitude, $\Gamma$ is the lead-dot hybridization strength.  From (a) to (f):  up 
and down spectral densities for the dot 1. 
Parameters are $\epsilon=-U/2$, $U=7$, and $\Gamma=1$. (g) Spin-spin correlation function for the P arrangement ($1$ and $2$ sub-indices denote the left and right dot, respectively). (h) Double dot phase diagram for the P case when $t$ and $p$ are varied: the three phases are the Kondo phase, the anti-ferromagnetic phase (AFM) and the molecular-orbital phase (MO) in which for a very large $t$ the two dots behave as an effective one-dot system with energy levels at $\epsilon\pm t$. In this case, the usual spin-$1/2$ Kondo physics emerges.  Lower inset: Double dot system in the AP configuration  [majority(minority) spins in the left reservoirs are down(up) spins and majority(minority) spins in the right reservoir are spin-up(down) electrons]. Up and down spectral  densities for the dot 1 in the AP lead orientation are plotted in (i),  (j), (k), and (l) . Parameters are $\epsilon=-U/2$, $U=7$, and $\Gamma=1$.}
\label{fig1}
\end{figure}

%Such systems are also interesting from the perspective of the dynamical mean-field theory (DMFT) studies of correlated electron systems in external magnetic fields: in cluster DMFT, such bulk models map to impurity models of the same class as discussed in this work. 

In this letter we investigate the role of injecting spin-polarized 
charge carriers in serially coupled QDs (see upper inset in Fig. 1). 
We report pronounced transport phenomena due to the coupling 
with ferromagnetic electrodes, which are not present in single dots. 
In the absence of the particle-hole (p-h) symmetry,  ferromagnetic
contacts induce effective magnetic field in each QD.
%, denoted by $B_{1(2) \mathrm{eff}}$. 
We show that for parallel magnetization in the electrodes,
these induced effective fields can be compensated by properly tuning  the inter-dot tunneling amplitude $t$,
thereby  restoring  the Kondo states  in each QD.  
Furthermore, we demonstrate the behaviour of the DD set-up as a controllable spin-filter device
by gating the dots and varying either the lead polarization or, which is more useful, the inter-dot tunnelling $t$.
Finally, we complete our study by analysing the role of the polarization on the 2IKP.
In particular, we find that the equivalent model with a ferromagnetic spin-spin interaction
$J_{\mathrm{FM}}$ behaves as a spin-$1$ exchange-anisotropic Kondo model and exhibits a Kosterlitz-Thouless
quantum phase transition at finite polarization.

{\it Model} -- The double dot set-up (Fig. 1) is modelled as a two-site
Hubbard model. The Hamiltonian reads:
\begin{multline}
\mathcal{H}= \sum_{k,\sigma} \varepsilon_{\alpha k\sigma} c_{\alpha k\sigma}^{\dagger} c_{\alpha k\sigma} 
+ \sum_{\sigma} \epsilon n_{\alpha\sigma} + U n_{\alpha\uparrow} n_{\alpha\downarrow} \\
 + t \sum_{\sigma} \left( d^\dagger_{1\sigma} d_{2\sigma} + \text{h.c.} \right)
 + \sum_{k,\sigma} \left( V_{\alpha} c_{\alpha
 k\sigma}^{\dagger} d_{\alpha\sigma}  + \text{h.c.} \right),
\end{multline}
Here $c_{\alpha k\sigma}$ annihilates an electron in the electrode $\alpha \in \{1,2\}$ with wave-vector $k$ and spin 
$\sigma=\uparrow,\downarrow$. 
Similarly, $d_{\alpha\sigma}$ destroys an electron with spin $\sigma$ on the dot $\alpha$.
Each dot is also connected to a contact with hybridization amplitude $V_{\alpha}$. 
Ferromagnetic contacts are described by employing spin-dependent tunnelling rates:  $\Gamma_{\alpha,\sigma}=\pi |V_{\alpha}|^2\nu_{\alpha\sigma}$, 
$\nu_{\alpha\sigma}$ being
the spin-dependent density of states at in the contacts. The polarizations are parametrized by $p_\alpha$,
defined through $\Gamma_{\alpha,\uparrow}=\Gamma (1+p_\alpha)/2$, 
$\Gamma_{\alpha,\downarrow}=\Gamma (1-p_\alpha)/2$; we assume the total hybridization strength 
$\Gamma$ to be constant and equal for both contacts. 
%A schematic diagram of the system is shown in Fig.~1a. 
In this work we will consider
the case of collinear polarization between the leads with parallel (P)
($p_1=p_2=p$) and anti-parallel (AP) alignment  ($p_1=-p_2=p$).
We solve $\mathcal{H}$ using  the numerical 
renormalization group technique (NRG) \cite{Wilson:75}.

\emph{Particle-hole symmetric case, $\delta=\epsilon+U/2=0$}. --
The dot-$1$ spectral
functions $A_{1\sigma}(\omega)$ for the P and AP alignments are shown in Fig. 1. 
For $p=0$ and small $t$ [Fig. 1(a)], the spectral functions  for both spin alignments 
show a single peak (the Kondo resonance)
pinned at the Fermi level.
As $t$ increases, one moves
from the regime dominated by the Kondo effect to a phase governed
by the antiferromagnetic (AFM) coupling between the spins [Figs. 1(d), (e), and (f)]. 
The AFM regime is evidenced by a double-peak DOS with peaks at $\omega\approx\pm J/2$
 where $J=J_\mathrm{AFM}=4t^2/U$ is the superexchange interaction.
For finite polarization, $p\neq 0$, each dot is predominantly affected by the 
polarization of the neighboring contact
and only indirectly (thus weakly) by the other contact, therefore P and AP arrangements
show similar behavior [cf. Fig. 1(a)-(f) and Fig. 1(i)-(l)].  For small $t$, in the P orientation, 
[Fig. 1(c)] the spectral weight at $E_F$ becomes spin-dependent, 
i.e.,  $A_{1\uparrow}(E_F)\neq A_{1\downarrow}(E_F)$
but there is no spin-splitting.
%due to the absence of charge fluctuations when $\delta=0$. 
The width of the Kondo peak reduces according to \cite{PhysRevLett.91.247202} 
$T_K(p) \approx \tilde{D} \exp \{ [-1/(\nu_{\uparrow} J_K
+ \nu_{\downarrow} J_K)][\mathrm{\tanh^{-1}} (p)/p] \}$ where $\nu J_K=8\Gamma/\pi U$
at the p-h symmetry point. Thus, by increasing $p$ [Fig. 1(e)], the Kondo temperature is lowered.
When $J_\mathrm{AFM} \gtrsim 2T_K(p) $, the Kondo effect is destroyed and 
the dot spins bind into a local singlet which is reflected in a split spectral density [Fig. 1(e)].
The correlation function $\langle \mathbf{S}_1\cdot \mathbf{S}_2\rangle$ 
which measures the dot spins alignment  is shown in Fig. 1(g) for increasing $p$ for $t=-0.769$.  
Accordingly, the spectral function exhibits a splitting of magnitude $J_{\mathrm{AFM}}$ [Fig. 1(e)]
signaling the formation of the AFM singlet state. 
This is the same type of cross-over that one
observes in the $p=0$ model as a function of $t$.
% The cross-over can be observed as a smooth variation of the zero-frequency limit of the spin-down spectral function which, as a function of $p$, first increases and then smoothly decreases. 
This result implies that there is a continuous cross-over line in the
 ($p$, $t$) plane separating the Kondo phase from
the AFM one.
%In the very large $t$ limit the interactions play no role and we recover the bonding/anti-bonding molecular physics. 
A schematic phase diagram summarizing this result is depicted in Fig.~1(h). 
\begin{figure}[htbp]
\includegraphics[clip,width=8cm]{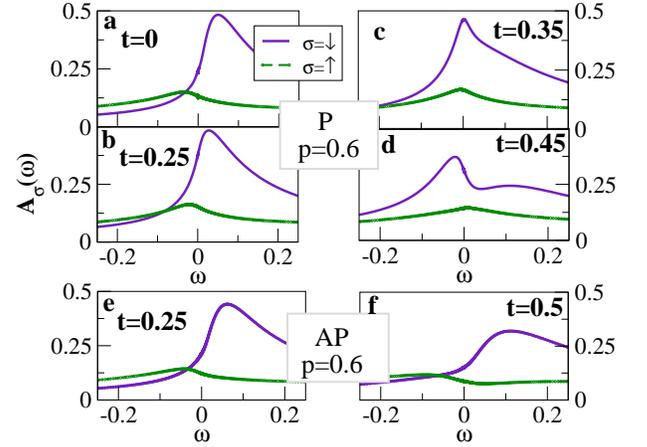}
\caption{(Color online) Spectral function of dot 1 when the
system is away from the particle-hole symmetric point.
Parameters are $U=7$, $\Gamma=1$, $\epsilon=-U/2+1$.}
\label{res3}
\end{figure}
%\begin{figure}[htbp]
%\includegraphics[clip,width=4cm]{diag.eps}
%%
%\caption{(Color online) Schematic phase diagram of the different
%regimes of the DQD system in the plane of tunneling coupling $t$
%and polarization $p$ for the case of parallel alignment of polarization
%in the contacts. AFM stands for the anti-ferromagnetic regime,
%and MO stands for molecular-orbital
%regime. The dashed line indicates the cross-over which is the remnant
%of the quantum phase transition in the exchange-only model.
%%Parameters as in Fig.~\ref{res1}.
%}
%%
%\label{diag}
%\end{figure}
Spectral dot densities in the AP orientation are shown in Figs. 1(i)-(l). As mentioned previously, 
they exhibit a similar behaviour than in the P configuration. We find remarkable differences 
between the two magnetic orientations (P) and (AP)  in the half-metallic limiting case when $p\to 1$  
[illustrated in Fig. 1(e), and (f) for the P case and Fig. 1(k), and (l) for the AP case]. For the P case  the 
spin-down spectral density at the Fermi level
$E_F$ ($\omega=0$ in the plots) vanishes, while for AP it remains finite. Conversely, for the P arrangement 
the spin-up spectral function is finite at $E_F$, while for AP it
goes to zero. The difference is due to an
interference effect that can be studied analytically
in the non-interacting ($U=0$) limit, with the result holding more generally.
For parallel alignment ($p\to 1$),  spin-down electrons are 
localized inside the DD system, while the spin-up electrons can 
still flow between the contacts; the $U=0$ spectral function for 
spin-down electrons has delta-like peaks at $\omega=\pm t$ 
and is zero elsewhere, while the spin-up spectral function is finite at $E_F$. 
For AP, in the 
$p=1$ limit the spin-up electrons from the contact $1$ can enter both dots, but cannot exit at 
the right contact. The $U=0$ spin-up spectral functions thus shows a peak at $E_F$ for the dot $2$ and, 
consequently, a null spectral density at $E_F$ for the dot $1$ due to interference. For the other electron spin, 
the behavior in the two dots is simply reversed. 
%(In the $J$-term model, both components of the spectral
%function go to zero for P and AP case alike). 
The difference in the $p \to 1$ limit becomes 
even more pronounced as $t$ increases.

\emph{General case, $\delta\neq0$}.  --
Away from the p-h symmetric point, the spin splitting is
generated by virtual processes proportional to the spin-dependent
hybridization functions: $B_{\mathrm{eff}}=\delta \epsilon_{\uparrow}
-\delta \epsilon_{\downarrow}$ (with $\bar\sigma=-\sigma)$
\begin{equation}
\delta \epsilon_{\sigma} \approx -\frac{1}{\pi}
\int d\omega
\left\{
\frac{\Gamma_{\sigma}(\omega)[1-f(\omega)]}
{\omega-\epsilon}
+
\frac{\Gamma_{\bar\sigma}(\omega)f(\omega)}
{\epsilon+U-\omega}
\right\}.
\end{equation}
Accordingly, at $p\neq0$ and for general values of parameters, the spectral 
function exhibits spin splitting (see Fig. 2). For small $t$, this splitting is fully analogous
to that observed in single QDs [e.g., see Fig. 2(a), and (b)]. For a single QD, in the AP case there is no 
induced field because of the direct compensation of the contributions from both contacts. In the DD
case, however, there is an induced field in each dot for both P and AP cases,
but they differ in the direction of the
fields: they are aligned along the same direction for the P case
($B_{1,\mathrm{eff}}=B_{2,\mathrm{eff}}$), while they point
in opposite directions for the AP case ($B_{1,\mathrm{eff}}=-B_{2,\mathrm{eff}}$). In single QDs, the splitting can only be
restored by the application of an external magnetic field. In
DDs, we find, in contrast, that the splitting compensation can also be  achieved by the exchange fields due to the inter-dot
exchange coupling $J_{\mathrm{AFM}}$: for a specific value of $t$ the compensation occurs [Fig. 2(c)] and beyond this 
$t$ value the splitting is shown again [Fig. 2(d)].
This can be understood as follows:
seen from dot  $1(2)$, the exchange coupling $J_{\mathrm{AFM}}$ between the dots can be regarded as an
effective magnetic field: $B_{\mathrm{exc}} = J_{\mathrm{AFM}} S_{z2(1)}$.
The restoration is, however, only
possible in the P case while in the AP case the splitting only grows
when $t$ is increased [see Fig. 2(e) and (f)]. Note that this type of restoration of the Kondo peak has also
been predicted in the two-impurity Kondo problem in an external magnetic field \cite{Simon:05,PhysRevLett.94.086805}
and verified experimentally \cite{PhysRevLett.96.017205}.
\begin{figure}[htbp]
\includegraphics[clip,width=8.5cm]{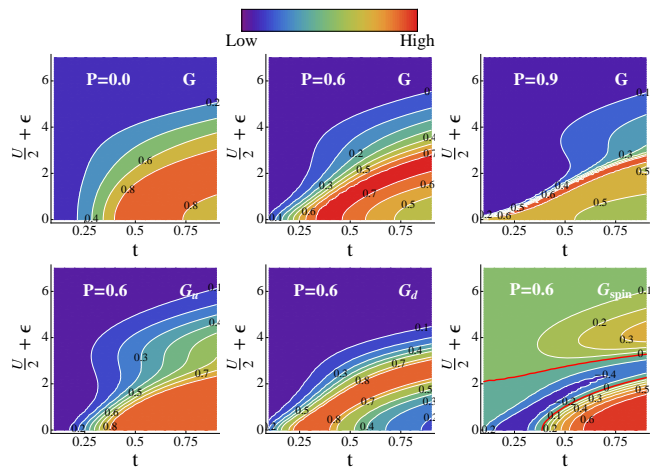}
\caption{(Color online) Conductance plots versus $\delta=\epsilon+U/2$ and $t$. Upper panel: 
Linear conductance  for $p=0,0.6,0.9$ in the parallel arrangement.
Lower panel: spin-up (left), spin-down (center) conductance and
spin conductance (right) for $p=0.6$. Parameters are $U=7$, $\Gamma=1$.}
\label{res2}
\end{figure}
Interestingly, all these features shown in the dot spectral density are reflected in a measurable transport
magnitude: the linear conductance. We compute the conductance through the DD system 
as a function of the inter-dot tunnelling
$t$ and the dot level position $\epsilon$ for
three values of $p$ in Fig.~3  (upper panel). 
For $p=0$ we observe the standard results for the
conductance of a DD: along
the particle-hole symmetric line the conductance is low in the small-$t$
and large-$t$ limits, and it peaks at the cross-over point $J_{\mathrm{AFM}} \sim
2 T_K$. For very large $t$, the conductance becomes high when
one of the molecular orbitals at energies $\epsilon \pm t$ is tuned to
the Fermi level signaling the molecular orbital phase. The cross-over
conductance peak and the molecular-orbital conductance peak are
smoothly connected in the ($t$,
$\epsilon$) plane by a high-conductance ridge which reaches 
the \textit{unitary conductance}, $G=2e^2/h$. The influence of polarized contacts is more
dramatic for the P alignment, due to the interplay between $J_{\mathrm{AFM}}$ and the effective fields.
The DD can namely behave as a \emph{spin-filter}
device for a  a finite polarization in the P arrangement. 
This can be seen from the spin-up conductance $G_{\uparrow}$,
the  spin-down conductance $G_{\downarrow}$, and 
the  spin conductance $G_s=G_{\uparrow}-G_{\downarrow}$,
which are shown in the  lower panel of Fig.~3.
For finite $p$ the conductance ridge no longer reaches the unitary limit.
The maximum conductance coincides with 
the restoration of the Kondo effect when $B_{\mathrm{exc}}+B_{\mathrm{eff}}=0$.
$G_{\downarrow}$ exhibits a sharp
peak in the ($t$, $\epsilon$) plane; the narrow width is due to the
weak hybridisation of the spin-down electrons 
%($\textcolor{NavyBlue}{\Gamma_{\downarrow}}$)
that reduces the Kondo scale as $p$ increases. 
% Of course, thus a sharp resonance criterion needs to be met for the \textcolor{NavyBlue}{spin-down}  
%electrons to traverse the system. Therefore,  \textcolor{NavyBlue}{$G_{\downarrow}$} is more diffuse, but it makes a smaller contribution to the total conductance. 
As a consequence,  one finds that \emph{the direction and amplitude of the spin 
conductance $G_s$ can be tuned by the parameters $\epsilon$ and $t$} (shown in Fig.~3).
Therefore the DD acts as an efficient spin-filter device with potential applications in the construction
of spintronic devices.

\textit{Equivalent two-impurity Kondo model}. --
We also study the associated Kondo model with
direct inter-dot magnetic exchange coupling of the form $\mathcal{H}=J_{}
\vc{S}_1 \cdot \vc{S}_2$ and no hopping term, i.e., the 
hopping term $t$ is replaced by a pure exchange term with $J=J_{\mathrm{AFM}}=4t^2/U$.
Provided that the charge transfer term between the leads 
(which corresponds to a cotunneling term through the two dots)
can be neglected, one finds a true quantum phase transition  which
corresponds to the critical behavior in the 2IKM \cite{Sela:09}.
We found that polarization in the leads does not affect
this conclusion. 
However, in an experimentally realistic system,  this phase transition is replaced by
a smooth cross-over as a function of $t$ \cite{Logan:11}.

For completeness, we have also analysed the case of a 
ferromagnetic exchange interaction, {\it i.e.}, $J=-J_{\mathrm{FM}}<0$.
The corresponding spectral functions are
shown in Fig.~4. For normal electrodes and small 
$J_{\mathrm{FM}}$, each dot exhibits a single-channel spin-$1/2$ Kondo state, while for a large 
$J_{\mathrm{FM}}$, the dot spins first rigidly bind  at some high temperature 
into a spin-$1$ state which is then screened at a lower temperature.
%screened in a two-stage procedure at lower temperatures (notice, that since we consider
%identical Kondo couplings for each dot the screening procedure occurs in a single stage). 
%Our NRG calculations for the impurity magnetic susceptibility, $\chi$ and 
% the entropy, $E$ evidence the formation of spin-$1$ Kondo state. 
Note that the shape of the Kondo resonance is therefore different for small and 
large-$J_{\mathrm{FM}}$ cases [corresponding to Fig.~4(a) and Fig. 4(b), respectively]. 
At finite $p$ [Fig.~4(c), and Fig.~4(f)], 
the Kondo effect  is destroyed and the NRG results show that the DD set-up has a residual $\ln 2$
entropy, indicating the occurrence of a quantum phase transition 
from the Kondo screened phase to a partially-screened phase. The
transition is of the Kosterlitz-Thouless type and it occurs at a value of $p$ which depends on $J_{\mathrm{FM}}$. Thus, there is a line of
phase transitions in the ($p$, $J_{\mathrm{FM}}$) parameter plane. This behavior is not
unexpected. It should be recalled that for a single QD, the
presence of spin polarization results in the exchange-anisotropy of the
effective low-energy Kondo model, $J_\perp \neq J_z$. The same effect
is expected in the present spin-$1$ Kondo model. It is known, however,
that anisotropic high-spin Kondo model feature partially-screened
states in their phase diagrams \cite{PhysRevB.77.075114}.

\begin{figure}[htbp]
\includegraphics[clip,width=8cm]{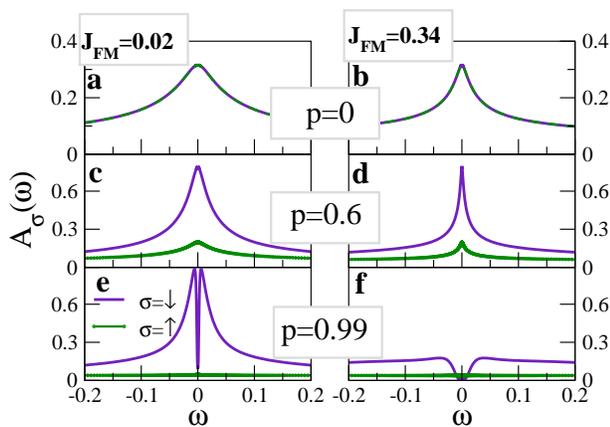}
\caption{(Color online) Spectral functions for the model with
ferromagnetic exchange coupling  for 
three different polarizations $p=0$, $p=0.6$ and $p=0.9$ (P case) and 
two distinct $J_{\mathrm{FM}}$ strength couplings.}
%Solid (green) lines correspond
%to the spin-up  dot 1 spectral function 
%and dashed (purple) lines show the spin-down
%dot spectral function.}
%
\label{fig4}
\end{figure}

{\it Conclusion}. -- Ferromagnetic contacts alter profoundly the transport properties
of serially coupled double dots. 
%For the particle-hole symmetry 
%the spin-polarized electrodes induce strong electron localization on the dots.
%Away from the particle-hole symmetry, 
The interplay
between the anti-ferromagnetic exchange coupling and the effective fields induced by the polarized leads 
in the parallel contact configuration may reinforce the Kondo effect. By
tuning the interdot tunnelling and the dot gates the spin conductance 
reverses its sign, thus the DD acts as a spin-filter device. We have also shown that the 2IKM with ferromagnetic 
interaction exhibits a Kosterlitz-Thouless type transition ascribed to the partially spin polarized contacts.
We propose double dot carbon nanotubes attached to polarized contacts
as the best candidate to 
observe our predictions.  In carbon nanotube quantum dots, large polarizations and 
much stronger Kondo states have been 
observed in comparison 
with semiconductor quantum dots \cite{PhysRevB.73.241401,0268-1242-21-11-S11,Lindelof,PhysRevB.80.035412}

\begin{acknowledgments}
RZ acknowledges the support of the Slovenian Research
Agency (ARRS) under Grant No. Z1-2058. J.S and R.L. are supported by Spanish MICINN (Grant No. FIS2008-00781)
and the Conselleria d’Innovació, Interior i Justicia Govern de les Illes Balears.
\end{acknowledgments}
\bibliographystyle{apsrev}
\bibliography{mybibliographydqd.bib}
\end{document}

\title{Tunable Kondo effect in double quantum dot coupled to ferromagnetic
contacts}
\author{Rok \v{Z}itko}
\affiliation{Jo\v{z}ef Stefan Institute, Jamova 39, SI-1000 Ljubljana,
Slovenia}
\affiliation{Faculty  of Mathematics and Physics, University of Ljubljana,
Jadranska 19, SI-1000 Ljubljana, Slovenia}
\author{Jong Soo Lim}
\affiliation{Departament de F\'{i}sica, Universitat de les Illes Balears,
  E-07122 Palma de Mallorca, Spain}
  \author{Rosa L{\'o}pez}
\affiliation{Departament de F\'{i}sica, Universitat de les Illes Balears,
  E-07122 Palma de Mallorca, Spain}
\affiliation{Institut de F\'{i}sica Interdisciplinar i de Sistemes Complexos
  IFISC (CSIC-UIB), E-07122 Palma de Mallorca, Spain}
\author{Jan Martinek}
\affiliation{Institute of Molecular Physics,
  Polish Academy of Sciences, Smoluchowskiego 17,
  60-179 Pozna\'n, Poland}
\author{Pascal Simon}
\affiliation{Laboratoire de Physique des Solides, CNRS UMR-8502, Univ. Paris Sud, 91405 Orsay Cedex, France}
%\date{\today}

%\begin{abstract}
%\end{abstract}

\maketitle

\newcommand{\expv}[1]{\langle #1 \rangle}
\newcommand{\vc}[1]{\mathbf{#1}}
\newcommand{\korr}[1]{\langle\langle #1 \rangle\rangle}

\section{Equation of motion study of the non-interacting case, $U=0$}

We consider the $U=0$ limit of the Hamiltonian presented in the main
text. Our goal is to compute the spectral functions using the
equation-of-motion method:
\begin{equation}
z \korr{A;B}_z = \expv{[A,B]_+}+\korr{[A,H]_-; B}_z
\end{equation}
where $z$ is the frequency parameter in the complex plane,
$\korr{A;B}_z$ denotes the correlator between the operators $A$ and
$B$, while $[A,B]_-$ and $[A,B]_+$ are a commutator and an
anticommutator, respectively. We introduce the notation
$G_{i\sigma,j\tau}(z)=\korr{ d_{i\sigma} ; d_{j\tau}^\dag}_z$ for the
impurity Green's functions, where $i,j \in \{1,2\}$ and $\sigma,\tau
\in \{\uparrow,\downarrow\}$. Since the spin components along the
$z$-axis are conserved, only $\sigma=\tau$ parts of the Green's
function are non-zero, thus we may also write $G_{ij,\sigma}(z) \equiv
G_{i\sigma,j\sigma}(z)$. Furthermore, we define $G_{\alpha k
i,\sigma}(z) = \korr{c_{\alpha k\sigma};d_{i\sigma}^\dag}_z$ where $k$
is a wave-vector in the lead $\alpha \in \{1,2\}$.

We obtain (omitting the argument $z$ of Green's functions for brevity)
\begin{equation}
\begin{split}
z G_{11,\sigma} &= 1 + \epsilon G_{11,\sigma} + t G_{21,\sigma} + V_1 \sum_k G_{1k1,\sigma},\\
z G_{12,\sigma} &=     \epsilon G_{12,\sigma} + t G_{22,\sigma} + V_1 \sum_k G_{1k2,\sigma},\\
z G_{21,\sigma} &=     \epsilon G_{21,\sigma} + t G_{11,\sigma} + V_2 \sum_k G_{2k1,\sigma},\\
z G_{22,\sigma} &= 1 + \epsilon G_{22,\sigma} + t G_{12,\sigma} + V_2 \sum_k G_{2k2,\sigma},
\end{split}
\end{equation}
and
\begin{equation}
z G_{\alpha k i,\sigma} = \epsilon_{\alpha k \sigma} G_{\alpha k i,\sigma} + V_\alpha G_{\alpha i,\sigma}.
\end{equation}
From the latter equation it follows
\begin{equation}
G_{\alpha k i,\sigma} = \frac{V_\alpha}{z-\epsilon_{\alpha k \sigma}} G_{\alpha i,\sigma}.
\end{equation}
We introduce the (spin-dependent) hybridization function as
\begin{equation}
\Delta_{\alpha\sigma} = \sum_k \frac{V_\alpha^2}{z-\epsilon_{\alpha k \sigma}}.
\end{equation}
The spin dependence arises from the $\sigma$-dependence of the band
dispersion $\epsilon_{\alpha k \sigma}$, i.e., from the spin
dependence of the conduction-band density of states.

Solving the system of equation we obtain
\begin{equation}
\begin{split}
G_{11,\sigma}(z) &= \frac{1}{z-\epsilon-\Delta_{1\sigma} -
\frac{t^2}{z-\epsilon-\Delta_{2\sigma}}}, \\
G_{22,\sigma}(z) &= \frac{1}{z-\epsilon-\Delta_{2\sigma} -
\frac{t^2}{z-\epsilon-\Delta_{1\sigma}}}.
\end{split}
\end{equation}
The spectral functions are then simply obtained as
\begin{equation}
A_{i\sigma}(\omega) = -\frac{1}{\pi} \mathrm{Im}
G_{ii,\sigma}(\omega+i\delta),
\end{equation}
where $\delta \to 0^+$.

We now consider the half-metallic limit, $p \to 1$.
For parallel alignment, this implies
\begin{equation}
\Delta_{1\uparrow}=\Delta_{2\uparrow}=\Delta, 
\quad\quad \Delta_{1\downarrow}=\Delta_{2\downarrow}=0.
\end{equation}
We take the wide-band limit where $\Delta=-i\Gamma$.
We obtain
\begin{equation}
A_{1\uparrow}(\omega)=A_{2\uparrow}(\omega)=\frac{\Gamma}{\pi}
\frac{(\omega-\epsilon)^2+t^2+\Gamma^2}
{\left[(\omega-\epsilon+t)^2+\Gamma^2 \right]
 \left[(\omega-\epsilon-t)^2+\Gamma^2 \right] },
\end{equation}
and
\begin{equation}
A_{1\downarrow}(\omega)=A_{2\downarrow}(\omega)=\frac{1}{2} \delta(\omega-\epsilon-t)
+\frac{1}{2} \delta(\omega-\epsilon+t).
\end{equation}

For antiparallel alignment we have
\begin{equation}
\Delta_{1\uparrow}=\Delta_{2\downarrow}=\Delta, \quad\quad
\Delta_{1\downarrow}=\Delta_{2\uparrow}=0.
\end{equation}
It follows
\begin{equation}
A_{1\uparrow}(\omega)=A_{2\downarrow}(\omega)=
\frac{\Gamma}{\pi} \frac{(\omega-\epsilon)^2}{
t^4+
2t^2(\omega-\epsilon)^2+
(\omega-\epsilon)^2 [\Gamma^2+(\omega-\epsilon)^2]
}
\end{equation}
and
\begin{equation}
A_{2\uparrow}(\omega)=A_{1\uparrow}(\omega)=
\frac{\Gamma}{\pi} \frac{t^2}{
t^4-
2t^2(\omega-\epsilon)^2+
(\omega-\epsilon)^2 [\Gamma^2+(\omega-\epsilon)^2]
}.
\end{equation}
Note that at the particle-hole symmetric point ($\epsilon=0$), the
spectral functions $A_{1\uparrow}$ has a zero while $A_{2\uparrow}$ is
finite at $\omega=0$. This result also holds in the interacting case
with $U \neq 0$.

%\begin{acknowledgments}
%\end{acknowledgments}

%\bibliographystyle{apsrev}
%\bibliography{mybibliographydqd.bib}